\documentclass[preprint,review,12pt]{elsarticle}
\usepackage{graphicx}
\usepackage{amsmath}
\usepackage{longtable,array}
\usepackage{lscape}
\usepackage{color}
\usepackage{epsfig}

\newcommand{\ai}{{\it ab initio}}

\newcommand{\cm}{cm$^{-1}$}

\newcommand{\oz}{$^{16}$O$_3$}

\newcommand{\nuu}{$\nu_1$}
\newcommand{\nuuu}{$\nu_3$}
\newcommand{\muu}{$\mu$m}

\def\a0{{$a_{\rm 0}$}}

\journal{Journal of Quantitative Spectroscopy \& Radiative Transfer}

\begin{document}
\begin{frontmatter}

\title{Potential energy surface, dipole moment surface and
        the intensity calculations for the 10 \muu,  5 \muu\ and 3 \muu\ bands 
        of ozone}

\author{Oleg L. Polyansky\thanks{e-email: o.polyansky@ucl.ac.uk}}
\address{Department of Physics and Astronomy, University College London,
Gower Street, London WC1E 6BT, United Kingdom;
Institute of Applied Physics, Russian Academy of Science,
Ulyanov Street 46, Nizhny Novgorod, Russia 603950}

\author{Nikolai F. Zobov, Irina I. Mizus  and Aleksandra A. Kyuberis}
\address{Institute of Applied Physics, Russian Academy of Science,
Ulyanov Street 46, Nizhny Novgorod, Russia 603950}

\author{Lorenzo Lodi, Jonathan Tennyson}
\address{Department of Physics and Astronomy, University College London,
Gower Street, London WC1E 6BT, United Kingdom}

\date{\today}

\begin{abstract}
Monitoring ozone concentrations in the Earth's atmosphere using
spectroscopic methods is a major activity which undertaken both from the ground and from space. However there are long-running
issues of consistency between measurements made at infrared (IR)
and ultraviolet (UV) wavelengths. In addition, key
O$_3$ IR bands at 10 \muu, 5 \muu\ and 3 \muu\ also yield results
which differ by a few percent when used for retrievals. These problems
stem from the underlying laboratory measurements of the line intensities.
Here we use quantum chemical techniques, first principles
electronic structure and variational nuclear-motion calculations, to address this problem. 
A new high-accuracy \ai\  dipole moment 
surface (DMS) is computed. Several spectroscopically-determined
 potential energy surfaces (PESs)  are constructed by
fitting to empirical energy levels in the region below 7000 \cm\ starting
from an \ai\ PES. 
Nuclear motion calculations using these new surfaces allow the
unambiguous determination of the  
intensities of 10 \muu\ band transitions, and the computation of
the intensities of 10 \muu\ and 5 \muu\ 
bands within their experimental
error. A decrease in intensities within the 3 \muu\  is predicted which
appears consistent with atmospheric retrievals. The PES and DMS form a suitable
starting point both for the computation of comprehensive ozone line lists
and for  future calculations of electronic transition
intensities.
\end{abstract}

\end{frontmatter}

\section{Introduction}
The ozone molecule, O$_3$, is an important constituent of the 
Earth's atmosphere. At high altitudes its ultraviolet (UV) 
absorption bands protect life
from deadly solar UV radiation, while at low altitude ozone represents
a dangerous, poisonous pollutant. 
Monitoring of ozone concentration in the Earth's atmosphere is thus a major and important activity \cite{99Solomo.O3}
which undertaken both from the ground and from space. Much of this monitoring
is based on the use of remote sensing and therefore relies
on the availability of reliable laboratory spectroscopic data.

Another potential use of ozone spectroscopy is provided by remote
sensing of other planets, particularly exoplanets. The use of spectra
of key molecules whose presence in the atmosphere of an exoplanet could
point towards the possible presence of life, so-called biomarkers, is
the subject of active discussion \cite{14FrLiLo}.
One of the most important biomarkers is the presence of methane
in an oxygen-rich atmosphere  \cite{93SaThCa.CH4}.
As diatomic oxygen does not
have a strong IR spectrum due to its symmetry, oxygen's first
derivative ozone takes on the role as an important biomarker \cite{13GrGeGo}
alongside the O$_2$ A band.

Ozone can be monitored at both ultraviolet (UV) and infrared (IR)
wavelengths.
There are extensive compilations of spectroscopic data on ozone in
general spectroscopic databases such as HITRAN \cite{jt691}, GEISA
\cite{jt636} and the UV/Vis+ Spectral data base \cite{17NoHaFa}, as
well as the specialist compilations for missions such as MIPAS
\cite{03FlPiCa} and the ozone-specific 
Spectroscopy and Molecular Properties  of Ozone (SMPO) database
\cite{smpo}. Recommended cross sections for UV absorption
by ozone are regularly reviewed \cite{03OrChxx.O3,16OrStTa.O3}
and display a reasonable measure of self-consistency \cite{03Orphal.O3}.
The same is not true of the IR transition intensities which are
neither consistent between IR or bands between the IR and UV
\cite{12SmDeBe.O3}.

For atmospheric retrievals and monitoring, sub-1\% accuracy in
intensity is highly desirable. However, the 3 to 4 \% inconsistency
between various experimental observations of intensities of the ozone
IR absorption lines represents the present state of knowledge.  As
described in detail by Smith {\it et al.} \cite{12SmDeBe.O3}, the
major inconsistency is between several measurements of the strong 10
\muu\ absorption IR bands, which differ by 4 \%. Some of these
measurements give atmospheric retrievals in line with the UV observation of the Hartley band at
254 nm and some of them do not.  The second important inconsistency,
between retrievals based on the 10 \muu\ bands and 5 \muu\ bands
intensities, is described by Janssen {\it et al.}  \cite{16JaBoJe.O3}.
Based on measurements due to Thomas {\it et al.} \cite{10ThVoDe.O3},
it was shown by Janssen {\it et al.}  that the concentration of ozone
determined by a retrieval based on data for the 10 \muu\ band is 2 to 3~\%\
different from that based on the 5 \muu\ band when using 
spectroscopic data from HITRAN 2012
\cite{jt557}.

Recently Drouin {\it et al.}  \cite{17DrCrYu.O3}
attempted to validate the 10 \muu\ intensities by the simultaneous
measurement of IR and microwave absorption line intensities of ozone. This study
confirmed the
HITRAN 2012  intensities for the 10 \muu\ band within 1.5 \%;
however, the accuracy was limited by the signal to noise ratio
of the IR data. As 10 \muu\ band consists  of two vibration bands, namely $\nu_1$ and
the much stronger $\nu_3$ band, the signal to noise limitations should mean that for $\nu_3$
band the discrepancy should be even less.

Finally there also appears to be a consistency problem with ozone
intensities in the 3 \muu\ region. For this band the HITRAN 2016
\cite{jt691} compilation is based on measurements due to Bouazza {\it
  et al} \cite{95BoMiBa.O3} and the SMPO database \cite{smpo}. However
analysis by Toon \cite{toonoz} suggests that here too there are
systematic differences with line intensities when compared with results
for other wavelengths, in this case in the region of 9\%.

The ozone molecule has been studied extensively both experimentally
and theoretically, see Ref. \cite{17TyKaTa.O3} and references therein.
In particular, an accurate ground state potential energy surface (PES)
of ozone was determined long ago by Tyuterev and coworkers
\cite{99TyKaTa.O3}. This PES was  subsequently further improved
\cite{ozpesconf}. More recently   Tyuterev {\it et al.}  \cite{17TyKaTa.O3}
calculated an
\ai\ dipole moment surface (DMS) and  used it to
compute intensities of the 10 \muu\ band lines belonging to the \nuu\ and
\nuuu\ band. More details of and comparisons with this work are given below.

In this paper we concentrate only on \oz. We present a new \ai\ and
several fitted PESs for the ground electronic state of ozone, as well
as a new \ai\ DMS constructed for the purpose of calculating the
intensities of the 10 \muu, 5 \muu\ and 3 \muu\ bands of ozone.
Section II presents our \ai\ PES and DMS calculations.  The \ai\ PES
is constructed mainly to be the starting point for obtaining a
spectroscopically-determined PES. Conversely, semi-empirical
adjustment of the DMS usually leads to a deterioration in the intensity
calculations \cite{jt156,jt573}. Thus we use our \ai\ DMS for all
intensity calculations.  Section III describes the fit of the ozone
PES to spectroscopic data; we also provide a comparison of the
resulting PES with the existing ones. Section IV describes the
intensity calculations which are compared to the experimental values
and previous theoretical results.  Section V gives our conclusions and
plans for further work.

\section{Calculation of the \ai\ PES and DMS}

A completely global, \ai, ozone PES was recently constructed
\cite{14TyKaCa.O3} which removed previous problems with a spurious
hump in the dissociation region; this means that there is an available
\oz\ PES which is well-characterized for every geometry.

For our purposes we need both an analytical \ai\ PES and the values of
the \ai\ energies at hundreds of geometries, as both are used in our
procedure to fit the PES to experimental data; see, for example,
Bubukina {\it et al.} \cite{Bubukina2011}.  This is because our fit
procedure uses both empirical energies and \ai\ points with a reduced
weight; including these points prevents the final PES from moving too far away
from the \ai\ surface \cite{03YuCaJe.PH3}. This is done in order to
both avoid ``holes'' in the resulting PES, and to be able to fit more
PES parameters than is possible when only a limited number of
vibrational band origins are known.

We calculated both the PES and DMS at the same level of quantum
chemical theory. Experience with calculations of the DMS for  CO$_2$
\cite{jt613}  shows
that, if the fundamental band origins of the molecule are predicted to better
than 1 \cm, one can expect the sub-percent accuracy in the intensity
calculations provided the same level
 of theory is used for the DMS calculation and very accurate nuclear-motion wave
functions are used from a spectroscopically-determined PES.

These considerations resulted in the choice of the aug-cc-pwcVQZ basis
set \cite{05Hattig} and of the multi-reference configuration interaction (MRCI) method
for both PES and DMS calculations. Specifically, we used the
full-valence complete active space, in which the 12 valence electrons
are free to populate all 18 valence orbitals; with this choice of
active space the electronic wave function comprises 8,029
configuration state functions at the complete active space
self-consistent field (CASSCF) and 109,785 ones at the MRCI level.  We
used the newer MRCI code implemented in the Molpro package
\cite{molpro} using the Celani-Werner internal contraction scheme
\cite{00CeWexx.methods,11ShKnWe.methods}.  
Calculations at this level of theory took about 6 hours per geometry
running on a single modern CPU. 
 
We performed \ai\ 
calculations for 4300 geometries, 2637 of which are 
below 7000 \cm\ with respect to the bottom of the potential well.
The  Molpro package \cite{molpro} was  used for all electronic structure 
calculations. 

We fitted the \ai\ energies to the following functional form
\begin{equation*} \label{main_func_form}
V(S_1,S_2,S_3)= V_0+\left[\sum_{i,j,k}{K_{ijk}S_1^iS_2^jS_3^k}]\exp[-b_1(\Delta{r_1^2}+\Delta{r_2^2})\right]+
\end{equation*}
\begin{equation*}                           
            +E_0[\exp(-2\alpha\Delta{r_1})-2\exp(-\alpha\Delta{r_1})]+
\end{equation*}
\begin{equation*}                           
            +E_0[\exp(-2\alpha\Delta{r_2})-2\exp(-\alpha\Delta{r_2})]+
\end{equation*}
\begin{equation}                           
            +\exp[-b_2(\theta-\theta_e)]
\end{equation}
\begin{equation} \label{valence_coord}
S_1=(r_1 +r_2)/2 -r_e, S_2=(r_1 -r_2)/2, S_3 = \cos{\theta} -\cos{\theta_e}, \Delta{r_i}=r_i-r_e
\end{equation}
$E_0$=11500 \cm, $\alpha=3.31, b_1=2.15, b_2=10, r_e=1.282$ \AA, 
$\theta_e=116.88$ degrees.

Using 50 constants a root mean square (rms)  for the \ai\ energies of 1 \cm\ was obtained.
Table 1 presents the $J =0$ energy levels calculated using the resulting \ai\ PES
and compares them to  experimental values.
 Details
of the nuclear motion calculations for rovibrational energy levels
are given in the next section.

The dipole moment points were calculated using the finite field
procedure.  We have demonstrated for both water \cite{jt509} and
carbon dioxide \cite{jt613} that even though the finite field method
requires more calculations, the resulting DMS is more accurate.

A polynomial functional form is used to represent the DMS: 
\begin{equation} \label{corr_func_form}
\mu_x(r_1, r_2, \theta) = \sum_{i,j,k}{K_{ijk}^xS_1^iS_2^j\tilde{S}_3^k}
\end{equation}
\begin{equation} \label{dip_func_form}
\mu_y(r_1, r_2, \theta) = \sum_{i,j,k}{K_{ijk}^yS_1^iS_2^j\tilde{S}_3^k}
\end{equation}
where the angular valence coordinates differs from that used for the potential and is given by:
\begin{equation} \label{valence_coord2}
   \tilde{S}_3 =\theta_e -\theta
\end{equation}
where the X axis bisects the valence angle, the Y axis is perpendicular to X, 
and the  Z axis, for which $\mu_z=0$, is perpendicular 
to the molecular plane.
The X and Y components of the dipole were fitted separately.
$\mu_x$ has the symmetry properties $\mu_x(r_1, r_2, \theta)=\mu_x(r_2, r_1, \theta)$; the indices
were therefore selected by $j$ = 0, 2, 4, . . . . For $\mu_y$, the symmetry property is
$\mu_x(r_1, r_2, \theta)=-\mu_x(r_2, r_1, \theta)$ and the indices were restricted
to $j$ = 1, 3, 5, . . . .

For $\mu_X$, 53 constants were fitted and 730 
\ai\ points were used. 
The root-mean-square 
(RMS) deviation of the fit was $\sigma = 2.64 \times 10^{-5}$ au. 
For $\mu_y$, 43 constants were fitted and 2404 \ai\ points were used. The RMS
deviation of the fit was almost identical, $\sigma = 2.65 \times 10^{-5}$ au. 
A Fortran file containing the DMS is presented in the supplementary material.

\section{Fitting the PES to experimental energy levels}
In many different cases, such as for recent calculation for water
\cite{jt714}, it has been shown that the accuracy of the predicted
intensities for computed absorption lines depends on the quality of
the wave functions used to represent the lower and upper states in the
corresponding transitions. Our \ai\ PES is not good enough if we want
to aim for intensity predictions accurate to better than 1~\%.  Thus
fitting of the PES to the empirical energy levels is necessary.
The \emph{ab initio} PES was then fitted to selected experimental
energy levels using an iterative procedure based on the method
developed by Yurchenko \emph{et al.} \cite{03YuCaJe.PH3}.

Studies on the water molecule \cite{jt714} have shown that a
significantly more accurate PES can be obtained if we limit ourselves
to a restricted set of empirical energy levels.  In particular for
water, the highest energy involved in the fitting procedure
\cite{jt714} was 15 000 \cm, whereas the levels known from
conventional spectroscopy extends to 26 000 \cm; indeed
multi-resonance spectra \cite{Grechko2008,Maksytenko20071} even reach
and go beyond dissociation \cite{jt494}.  If we do not aim at
completeness or to cover all the available experimental data and
concentrate on the lower energies, significantly more accurate results
can be obtained. In particular, for H$_2$$^{16}$O when we fitted the
data up to 26 000 \cm\ \cite{Bubukina2011}, the standard deviation of
the fit was about 0.025 \cm. While limiting energies to below 15 000
\cm, more than a twofold improvement in accuracy was achieved with a
standard deviation of 0.011 \cm\ \cite{jt714}. Here we opted to fit
data significantly higher than the energies of the bands we are
interested in, but lower than the highest experimentally known energy
levels.

\subsection{Nuclear motion calculations}\label{fit_pars}

The nuclear-motion Schr\"odinger equation was solved using program
DVR3D \cite{jt338}, which makes use of an exact kinetic energy
operator. These calculations were performed in Radau coordinates and
used Morse-like oscillators \cite{jt14} with the values of parameters
$r_e\,=\,2.8$, $D_e\,=\,0.1$ and $\omega_e\,=\,0.0024$ in atomic units
for both radial coordinates, and associated Legendre functions for the
angular coordinate as basis functions.  The corresponding DVR grids
contained 20, 20 and 70 points for these coordinates, respectively.
The final diagonalized vibrational matrices had dimension 1500. For
the rotational problem, the dimensions of the final matrices were
obtained using the expression $400(J + 1 - p)$, where $J$ is the total
angular momentum quantum number and $p$ is the value of parity.
Atomic masses equal to 15.994915\,Da were used for oxygen; using
atomic rather than nuclear masses very approximately accounts for
non-adiabatic effects \cite{04Watson.diatom}.
 
\subsection{Optimization results}\label{res}

Empirical
$J$=0 energy levels were taken from the table in SMPO data base \cite{smpo},
$J$=2 and $J$=5 levels were obtained from HITRAN frequencies. This gave
a total of 371 levels with energies up to  $6000\,\text{cm}^{-1}$.
We varied the values of 36 
non-zero potential parameters of the starting PES, henceforth
PES ai, to obtain best agreement
with three separate sets of experimental energy levels;
as a result, we obtained three new 
potentials: 
\begin{enumerate}
 \item[PES1]  reproduces a set of 301 (70 were excluded from the fit)
empirical energy levels 
with a standard deviation of 0.027$\,\text{cm}^{-1}$. 
\item[PES2] reproduces a set of 350 (21 excluded) empirical energy levels
with a standard deviation of 0.225~$\text{cm}^{-1}$. 
\item[PES3]   reproduces the complete set of 371 empirical energy levels
with a standard deviation of 0.693 \cm. This rms is
dominated by the residues for so-called dark states which are only
reproduced with an rms of 1.6 \cm. The other levels have an rms 
of 0.3 \cm.
\end{enumerate}
Most of the excluded energy levels are highly excited or have high 
values of bending quantum number $\nu_2$; these levels are not well-described 
within our method which concentrated on getting a very accurate representation
of the stretching motions.

The method by Yurchenko \emph{et al.} \cite{03YuCaJe.PH3} allows us to
optimize simultaneously with respect to both the experimental rovibrational energy
levels and \emph{ab initio} energies. This avoids creating nonphysical
features in the optimized PES. For this purpose we used a set of 2637
\emph{ab initio} electronic energies in the energy region up to
$7000\,\text{cm}^{-1}$ (about 2.4\% of \emph{ab initio} points from
the original set of 2701 energies were excluded from the fit).
Standard deviations of our final PESs from this set of \emph{ab
  initio} data are about 69$\,\text{cm}^{-1}$,
87.2$\,\text{cm}^{-1}$, and 75.1$\,\text{cm}^{-1}$, respectively. At
the final stages of our fits the weights of \emph{ab initio} points
were $3\times10^{-9}$ for PES1 and 10$^{-8}$ for PES2 and PES3, with the
empirical data unit weighted.  This was sufficient to ensure the
physically correct behavior of the fitted potentials.

\begin{longtable}{rrrrrrrr}
\caption{Observed band origins, in \cm, of $^{16}$O$_3$ used in the
 fitting procedures and differences between observed and calculated 
 values (in cm$^{-1}$) for three new 
spectroscopically-determined potentials.}\\
\hline 
$\nu_1$ & $\nu_2$ & $\nu_3$ & $E_{\text{obs}}$ & PES ai & PES1 & PES2& PES3\\
\hline 
0 & 1 & 0 &  700.9310 & 4.43 & 0.0174 & 0.0570 & 0.0627 \\
1 & 0 & 0 & 1103.1370 & -1.86 & 0.0160 & 0.0249 & 0.0291 \\
0 & 2 & 0 & 1399.2730 &  8.70 & -0.0092 & -0.0120 & -0.0330 \\
1 & 1 & 0 & 1796.2620 &  2.51 & 0.0171 & -0.0153 & -0.0295 \\
0 & 0 & 2 & 2057.8910 & -9.38 & 0.0666 & 0.0453 & 0.0387 \\
0 & 3 & 0 & 2094.9920 & 12.81 & -0.0562 & -0.0588 & -0.1328 \\
2 & 0 & 0 & 2201.1550 & -3.92 & -0.0103 & 0.0146 & 0.0347 \\
1 & 2 & 0 & 2486.5760 &  6.72 & -0.0499 & -0.1087 & -0.1443 \\
0 & 1 & 2 & 2726.1070 & -5.13 & 0.0580 & 0.0983 & 0.1096 \\
0 & 4 & 0 & 2787.9000 & 16.66 &  & 0.5974 & 0.4494 \\
2 & 1 & 0 & 2886.1780 &  0.41 & -0.0228 & -0.0089 & -0.0205 \\
1 & 0 & 2 & 3083.7030 & -9.80 & 0.0036 & 0.1731 & 0.1679 \\
1 & 3 & 0 & 3173.9290 & 10.77 &  & 0.4359 & 0.3818 \\
3 & 0 & 0 & 3289.9300 & -5.63 & 0.0073 & 0.0254 & 0.0252 \\
0 & 2 & 2 & 3390.9180 & -0.97 & 0.0346 & 0.0494 & 0.0535 \\
0 & 5 & 0 & 3478.4000 & 20.73 &  & 0.0972 & -0.0892 \\
2 & 2 & 0 & 3568.0700 &  9.42 &  & 0.2299 & 0.2006 \\
1 & 1 & 2 & 3739.4270 & -5.63 & -0.0137 & 0.0236 & -0.0082 \\
1 & 4 & 0 & 3858.6000 & 15.19 &  &  & -1.7993 \\
3 & 1 & 0 & 3966.7000 & -1.43 & 0.0021 & 0.0799 & 0.0479 \\
0 & 0 & 4 & 4001.3140 & -13.92 & 0.0155 & -0.0096 & -0.0403 \\
2 & 0 & 2 & 4141.4180 & -0.06 & 0.0271 & -0.0015 & -0.0045 \\
2 & 3 & 0 & 4246.7000 &  8.84 &  & -1.0634 & -1.1265 \\
4 & 0 & 0 & 4370.3000 & -6.76 &  & -0.4440 & -0.5366 \\
1 & 2 & 2 & 4390.5000 & -1.43 &  & -0.1839 & -0.2234 \\
0 & 1 & 4 & 4632.8880 & -2.34 &  & -0.0333 & -0.1013 \\
3 & 2 & 0 & 4643.8000 & -4.54 &  & 0.0761 & 0.0251 \\
2 & 1 & 2 & 4783.4610 & -9.25 & -0.0460 & -0.0089 & -0.0272 \\
1 & 0 & 4 & 4922.5720 & -11.67 &  & 0.1503 & 0.2291 \\
4 & 1 & 0 & 5038.5000 & -3.08 &  & -1.2119 & -1.2083 \\
3 & 0 & 2 & 5172.0000 & -13.28 & -0.0653 & -0.4313 & -0.5617 \\
0 & 2 & 4 & 5266.9000 & -3.06 &  & -0.7619 & -0.8423 \\
3 & 3 & 0 & 5310.5000 &  5.94 & -0.1129 & 0.4639 & 0.6248 \\
5 & 0 & 0 & 5443.0000 & -6.07 &  &  & -3.0049 \\
1 & 1 & 4 & 5540.8980 & -6.23 &  &  & 1.6853 \\
4 & 2 & 0 & 5701.6000 &  0.48 &  & -0.0513 & -0.1379 \\
2 & 0 & 4 & 5766.3200 & -9.90 &  & 0.8783 & 1.3582 \\
3 & 1 & 2 & 5812.6000 & -9.55 &  & 0.3437 & 0.3426 \\
0 & 0 & 6 & 5997.0000 & -3.61 &  &  & -2.8861 \\
5 & 1 & 0 & 6100.2100 & -4.59 &  & -0.4007 & -0.5910 \\
0 & 0 & 1 & 1042.0840 & -5.10 & -0.0082 & -0.0696 & -0.0794 \\
0 & 1 & 1 & 1726.5220 & -0.76 & 0.0054 & -0.0051 & 0.0120 \\
1 & 0 & 1 & 2110.7840 & -6.45 & 0.0256 & 0.1036 & 0.1026 \\
0 & 2 & 1 & 2407.9350 &  3.41 & -0.0326 & -0.0952 & -0.0851 \\
1 & 1 & 1 & 2785.2390 & -2.25 & 0.0387 & 0.0256 & 0.0224 \\
0 & 0 & 3 & 3046.0880 & -12.59 & 0.0186 & 0.0639 & 0.0466 \\
0 & 3 & 1 & 3086.2180 &  7.43 & -0.0419 & -0.1323 & -0.1696 \\
2 & 0 & 1 & 3186.4110 & -8.92 & 0.0261 & 0.0822 & 0.0687 \\
1 & 2 & 1 & 3455.8240 &  1.78 & -0.0313 & -0.1369 & -0.1418 \\
0 & 1 & 3 & 3698.2920 & -8.33 & -0.0006 & 0.0951 & 0.0638 \\
2 & 1 & 1 & 3849.9110 & -4.81 & -0.0066 & 0.0304 & 0.0207 \\
1 & 0 & 3 & 4021.8500 & -11.70 & 0.0099 & 0.0544 & 0.0387 \\
1 & 3 & 1 & 4122.0690 &  5.61 &  & 0.2899 & 0.2788 \\
3 & 0 & 1 & 4250.2230 & -10.27 & -0.0069 & -0.0404 & -0.1484 \\
0 & 2 & 3 & 4346.7270 & -4.03 & 0.0997 & 0.2117 & 0.1666 \\
2 & 2 & 1 & 4508.1320 & -0.99 &  & 0.1729 & 0.1588 \\
1 & 1 & 3 & 4658.9500 & -7.15 & 0.0201 & -0.1270 & -0.1879 \\
1 & 4 & 1 & 4783.2000 &  9.21 &  & -0.9981 & -0.9682 \\
3 & 1 & 1 & 4897.2770 &  2.43 & 0.0222 & -0.0288 & -0.0254 \\
0 & 0 & 5 & 4919.2030 & -21.86 &  & 0.0359 & 0.0371 \\
2 & 0 & 3 & 5077.0950 & -4.79 & -0.0087 & -0.3957 & -0.3887 \\
1 & 2 & 3 & 5291.1710 & -2.47 & 0.0076 & 0.0836 & 0.0483 \\
4 & 0 & 1 & 5307.7900 & -11.83 & 0.0217 & -0.4022 & -0.6822 \\
0 & 1 & 5 & 5518.8120 & -6.70 &  & 0.0990 & 0.2440 \\
3 & 2 & 1 & 5562.0000 & -1.18 &  &  & -2.8649 \\
2 & 1 & 3 & 5697.3230 & -12.88 &  & -0.0054 & 0.0216 \\
1 & 0 & 5 & 5783.7850 & -9.48 &  &  & 1.7396 \\
4 & 1 & 1 & 5947.0700 & -8.37 &  & -0.7589 & -0.8447 \\
1 & 0 & 5 & 6063.9220 & -5.57 &  & -1.1198 & -1.2258 \\
\hline
\end{longtable}

Table 1 presents the values of the $J=0$ band origins
calculated using  the ab initio PES, PES1, PES2  and PES3.  
Fortran files representing all
the PESs are provided as part of
the supplementary material. Figure 1 shows a cuts through the potential (PES1) and \ai\ DMS.

\begin{figure}
\centering
\scalebox{0.6}{\includegraphics[scale=0.8]{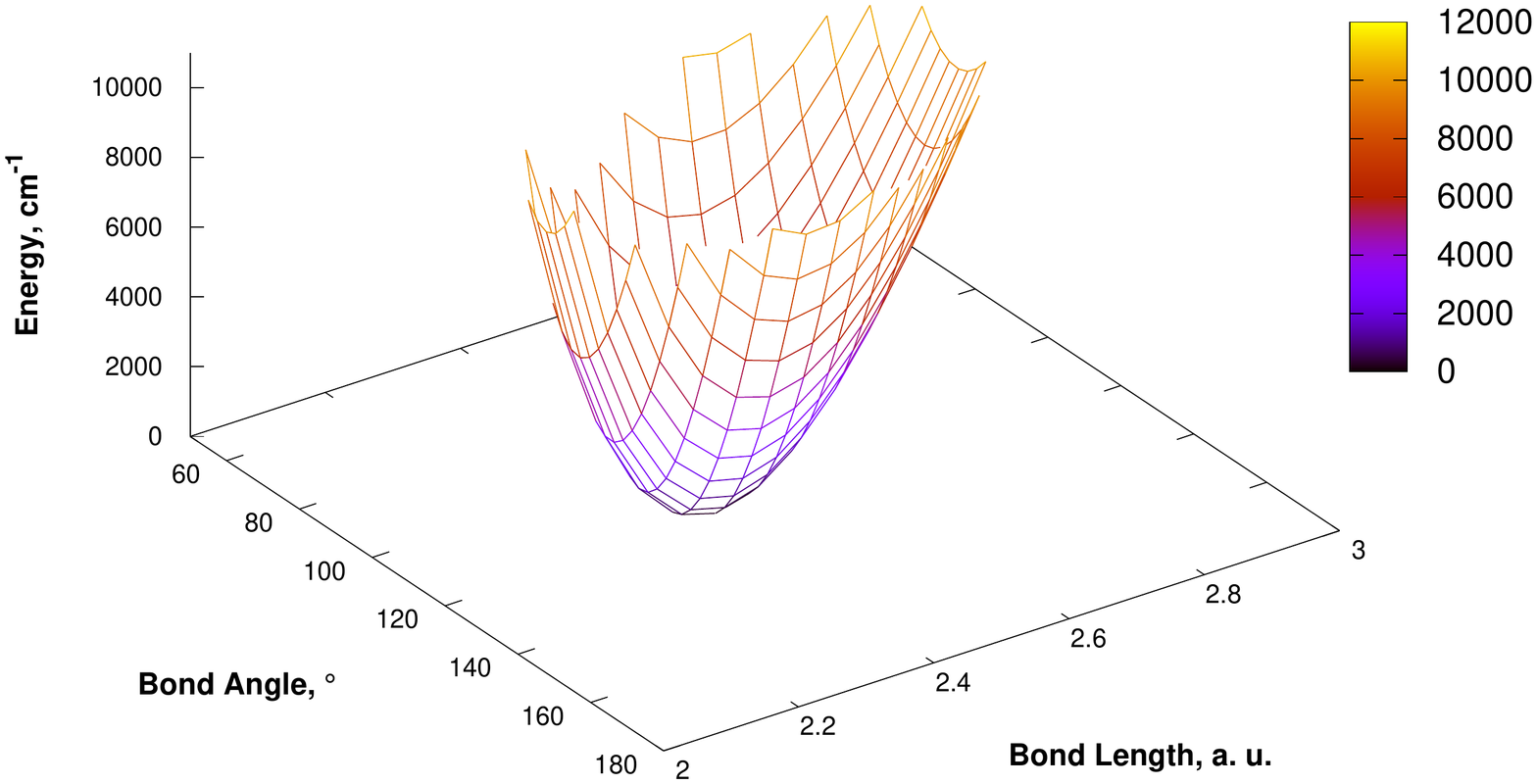}}
\scalebox{0.6}{\includegraphics[scale=0.8]{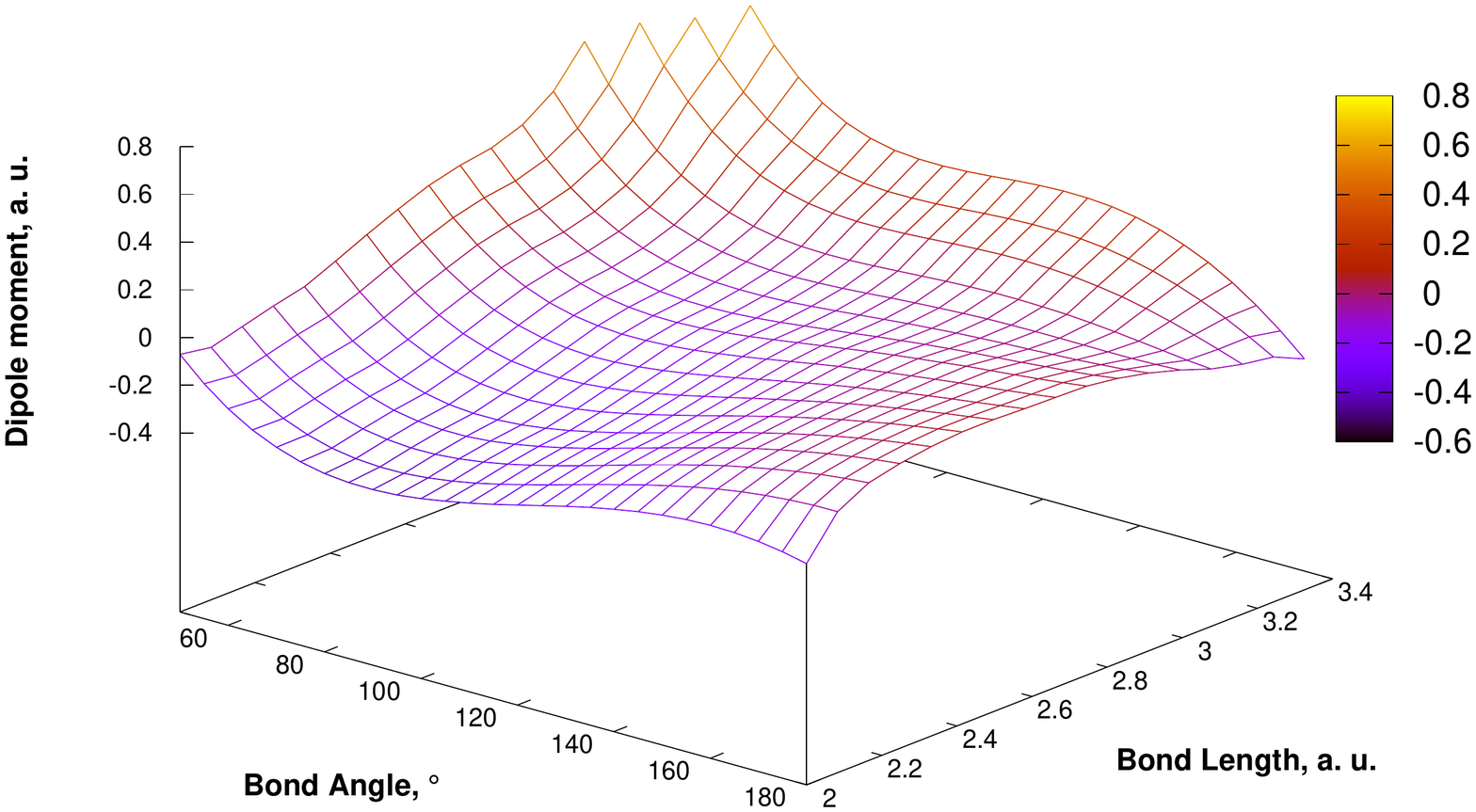}}
\caption{Cuts through the PES1 potential (upper) and dipole moment surface (lower). Plots are
for isosceles  triangle geometries with both bondlengths kept the same. The DMS plot gives
the ``bisector'' z-component of the dipole as for isosceles geometries the other components of the dipole are zero.}
\label{f:1}
\end{figure}

\section{Intensity of ozone absorption lines in the 10 \muu, 5 \muu\ and 3 \muu\  bands}

All absolute intensities quoted in are in \lq\lq HITRAN units'' of
cm$^2$/(molecule cm) and converted to 100 \% \oz.

In order to calculate
line intensities analytical forms for both the PES and DMS must
be provided to the DVR3D program suite. Comparisons were made for
various PESs and DMSs: our \ai\ DMS, PES1 and surfaces from the 
recent work of Tyuterev {\it et al.} \cite{17TyKaTa.O3}.

Table 2 and 3 present comparisons of calculated intensities with
empirical data used in  HITRAN 2016 
and which are taken from the SMPO \cite{smpo}
database. 
Table 2 presents a comparison of the calculated intensities using DMS
F2 of Tyuterev {\it et al.} \cite{17TyKaTa.O3} and our DMS using the
wave functions produced by our PES1 for the 10~\muu\ (001) band which
is ozone's strongest IR band.  Comparison of the results shows that
use of DMS F2 gives results that differ from those given by our DMS by
almost 4 \%\ when the same PES is used. We do not have access to the
PES used by Tyuterev {\it et al.} \cite{17TyKaTa.O3}, hence we cannot
present the results using their PES and our DMS. The results using
their PES and DMS are therefore taken from the table X of their paper
\cite{17TyKaTa.O3}. The difference with our results (last column) is
slightly lower -- only 3 \%. It would seem that the difference of 4\%\
between different DMSs is compensated by about 1 \% due to the use of
different PESs. The discrepancy with experiment given by our PES1
and DMS is significantly lower than other combinations: less than 1 \%.

Table 3 gives a similar comparison for the much weaker (100) band at
9~\muu. The difference between calculations performed with Tyuterev
{\it et al.}'s DMS F2 and our DMS is the same 4 \%.  However, for this
band the discrepancy with HITRAN 2016 is slightly worse in our
calculations.

Results for the 5 \muu\ (101) band line intensities are presented in
the Table 4. One can see that as for the strong (001) 10 \muu\ band,
that our 5 \muu\ band predicted intensities agree with the observed
intensities within the experimental uncertainties \cite{10ThVoDe.O3}.

The last region of interest for the problem of retrieval
inconsistencies is 2900 \cm\ -- 3000 \cm\ region. As reported by Toon
\cite{toonoz} ``The 2900-3000 \cm\ also produce 8-9\% too low \oz\
amounts in all line lists.'' This means that intensities of the
experimental lines presented in these line lists are too large by about
9\%. Table 5 presents intensities for the (003) band. It can be seen
that our calculated intensities are between 7 and 11 \% lower than the
HITRAN values.  This indicates that our calculated intensities should
solve the 3000 \cm\ region inconsistency problem as well.
We note that our results show a distinct difference between P-branch
transitions  (difference relative to HITRAN about 7\%) and R-branch
ones (difference 11\%). At this stage it is unclear if this issue
is associated with the measurements or our calculations.

\begin{longtable}{rrrccc}
\caption{Intensities of the \oz\ \nuuu\ band  calculated using PESs and DMSs
from Tyuterev {\it et al.}\cite{17TyKaTa.O3} and this work.
HITRAN intensities are given as 100\%\ abundance in units of cm/molecule
with powers of 10 in parenthesis. Other intensities are given
as a percentage difference to these values.}\\
\hline
Assignment& $\tilde{\nu}$ &HITRAN &(o-c)/c\%                &(o-c)/c\%                &(o-c)/c\%  \\
 (001)    &  (\cm)  &    &DMS F2 \cite{17TyKaTa.O3}&DMS F2 \cite{17TyKaTa.O3}&DMS(this work)\\ 
          &    &         &PES \cite{17TyKaTa.O3}   &PES1(this work)          &PES1(this work)\\ 
\hline
15 0 14 0 &   1052.848  & 4.09($-$20)   &  -2.7     & -3.6  & 0.8    \\
17 0 16 0 &   1053.966  & 4.07($-$20)   &  -2.7     & -3.7  & 0.8    \\
16 1 15 1 &   1053.168  & 4.05($-$20)   &  -2.8     & -3.7  & 0.8    \\
15 1 14 1 &   1053.692  & 4.02($-$20)   &  -2.5     & -3.5  & 0.9    \\
14 1 13 1 &   1051.985  & 3.99($-$20)   &  -2.7     & -3.6  & 0.8    \\
17 1 16 1 &   1055.006  & 3.99($-$20)   &  -2.6     & -3.5  & 0.9    \\
18 1 17 1 &   1054.289  & 3.98($-$20)   &  -2.8     & -3.7  & 0.8    \\
13 0 12 0 &   1051.657  & 3.97($-$20)   &  -2.6     & -3.6  & 0.8    \\
15 0 16 0 &   1027.456  & 3.97($-$20)   &  -2.4     & -3.3  & 1.0    \\
14 1 15 1 &   1028.495  & 3.94($-$20)   &  -2.4     & -3.3  & 1.1    \\
13 0 14 0 &   1029.433  & 3.94($-$20)   &  -2.4     & -3.3  & 1.0    \\
19 0 18 0 &   1055.016	& 3.93($-$20)   &  -2.8     & -3.7  & 0.8   \\     
16 1 17 1 &   1026.476	& 3.92($-$20)   &  -2.4     & -3.2  & 1.1   \\
13 1 12 1 &   1052.308	& 3.91($-$20)   &  -2.5     & -3.5  & 0.9  \\
17 0 18 0 &   1025.426	& 3.89($-$20)   &  -2.4     & -3.3  & 1.1  \\
19 1 18 1 &   1056.244	& 3.84($-$20)   &  -2.7     & -3.6  & 0.9  \\
12 1 13 1 &   1030.463	& 3.83($-$20)   &  -2.3     & -3.3  & 1.0  \\
16 2 15 2 &   1053.680	& 3.81($-$20)   &  -2.7     & -3.6  & 0.8  \\
15 2 14 2 &   1053.520	& 3.81($-$20)   &  -2.6     & -3.5  & 0.9  \\
17 2 16 2 &   1054.968	& 3.81($-$20)   &  -2.6     & -3.5  & 0.9  \\
20 1 19 1 &   1055.350	& 3.80($-$20)   &  -2.9     & -3.7  & 0.8  \\
15 1 16 1 &   1027.103	& 3.79($-$20)   &  -2.6     & -3.5  & 0.9  \\
12 1 11 1 &   1050.741	& 3.79($-$20)   &  -2.6     & -3.6  & 0.8  \\
18 1 19 1 &   1024.406	& 3.79($-$20)   &  -2.3     & -3.6  & 0.7  \\
13 1 14 1 &   1029.095  & 3.78($-$20)   &  -2.6     & -3.5  & 0.9  \\
11 0 12 0 &   1031.360  & 3.76($-$20)   &  -2.4     & -2.2  & 1.0  \\
14 2 13 2 &   1052.392  & 3.75($-$20)   &  -2.7     & -3.6  & 0.8  \\
18 2 17 2 &   1054.911  & 3.75($-$20)   &  -2.8     & -3.6  & 0.9  \\
11 0 10 0 &   1050.385  & 3.71($-$20)   &  -2.6     & -3.6  & 0.8  \\
19 0 20 0 &   1023.342  & 3.70($-$20)   &  -2.4     & -3.2  & 1.1  \\
21 0 20 0 &   1056.007  & 3.69($-$20)   &  -2.9     & -3.7  & 0.8  \\
17 1 18 1 &   1025.073  & 3.69($-$20)   &  -2.6     & -3.4  & 0.9  \\
19 2 18 2 &   1056.376  & 3.69($-$20)   &  -2.7     & -3.5  & 0.9  \\
13 2 12 2 &   1052.043  & 3.68($-$20)   &  -2.5     & -3.5  & 0.9  \\
14 2 15 2 &   1028.139  & 3.65($-$20)   &  -2.4     & -3.4  & 1.0  \\
11 1 10 1 &   1050.863  & 3.64($-$20)   &  -2.4     & -3.5  & 0.9  \\
16 2 17 2 &   1026.120  & 3.62($-$20)   &  -2.5     & -3.4  & 1.0  \\
11 1 12 1 &   1031.051  & 3.62($-$20)   &  -2.5     & -3.5  & 0.9  \\
15 2 16 2 &   1026.979  & 3.62($-$20)   &  -2.6     & -3.4  & 0.9  \\
21 1 20 1 &   1057.397  & 3.60($-$20)   &  -2.7     & -3.6  & 0.9  \\
13 2 14 2 &   1029.002  & 3.59($-$20)   &  -2.5     & -3.5  & 0.9  \\
20 2 19 2 &   1056.081  & 3.58($-$20)   &  -2.9     & -3.6  & 0.8  \\
10 1 11 1 &   1032.381  & 3.57($-$20)   &  -2.3     & -3.3  & 1.0  \\
20 1 21 1 &   1022.285  & 3.56($-$20)   &  -2.4     & -3.9  & 0.9  \\
12 2 11 2 &   1051.047  & 3.55($-$20)   &  -2.6     & -3.6  & 0.9  \\
12 2 13 2 &   1030.115  & 3.54($-$20)   &  -2.4     & -3.4  & 1.0  \\
22 1 21 1 &   1056.351  & 3.54($-$20)   &  -3.0     & -3.7  & 0.7  \\
17 2 18 2 &   1024.922  & 3.52($-$20)   &  -2.6     & -3.5  & 0.9  \\
18 2 19 2 &   1024.056  & 3.49($-$20)   &  -2.6     & -3.3  & 1.0  \\
19 1 20 1 &   1023.003  & 3.49($-$20)   &  -2.7     & -3.4  & 0.9  \\
\hline\hline
\end{longtable}
\newpage

\begin{longtable}{rrrccc}
\caption{Intensities of the \oz\ \nuu\ band  calculated using PESs and DMSs
from Tyuterev {\it et al.}\cite{17TyKaTa.O3} and this work.
HITRAN intensities are given as 100\%\ abundance in units of cm/molecule
with powers of 10 in parenthesis. Other intensities are given
as a percentage difference to these values.}\\
\hline
Assignment& $\tilde{\nu}$  & HITRAN  &(o-c)/c\%                 &(o-c)/c\%                 &(o-c)/c\%  \\
(100)     &  (\cm)      &      &DMS F2 \cite{17TyKaTa.O3} &DMS F2 \cite{17TyKaTa.O3} &DMS(this work) \\ 
          &             &            &PES \cite{17TyKaTa.O3}    &PES1(this work)           &PES1(this work)\\ 
\hline
25 1 24 0 &  1123.946  & 2.04($-$21)   &  2.3      & 1.4 & 5.4     \\
26 0 25 1 &  1124.295  & 2.03($-$21)   &  2.1      & 1.3 & 5.3     \\
24 0 23 1 &  1122.544  & 2.01($-$21)   &  2.4      & 0.8 & 6.4     \\
27 1 26 0 &  1125.524  & 2.01($-$21)   &  2.0      & 1.2 & 5.0     \\
23 1 22 0 &  1122.401  & 1.99($-$21)   &  2.6      & 1.7 & 5.6     \\
28 0 27 1 &  1126.022  & 1.98($-$21)   &  1.9      & 1.1 & 5.1     \\
22 0 21 1 &  1120.763  & 1.95($-$21)   &  2.7      & 1.8 & 5.7     \\
29 1 28 0 &  1127.129  & 1.93($-$21)   &  1.8      & 1.1 & 5.1     \\
21 1 20 0 &  1120.899  & 1.88($-$21)   &  2.9      & 2.0 & 5.9     \\
30 0 29 1 &  1127.731  & 1.87($-$21)   &  1.6      & 1.1 & 5.0     \\
20 0 19 1 &  1118.945  & 1.80($-$21)   &  3.0      & 2.1 & 6.0     \\
31 1 30 0 &  1128.752  & 1.79($-$21)   &  1.5      &     & 4.9     \\ 
32 0 31 1 &  1129.426  & 1.72($-$21)   &  1.4      &     & 5.7     \\ 
19 1 18 0 &  1119.446  & 1.70($-$21)   &  3.2      & 2.3 & 6.2     \\ 
18 0 17 1 &  1117.083  & 1.60($-$21)   &  3.3      & 2.4 & 6.3     \\ 
32 0 31 1 &  1129.426  & 1.72($-$21)   &  1.2      & 1.7 & 5.7     \\ 
34 0 33 1 &  1131.110  & 1.54($-$21)   &  3.6      &     & 4.6     \\ 
17 1 16 0 &  1118.049  & 1.48($-$21)   &  1.1      & 2.7 & 6.5     \\ 
35 1 34 0 &  1132.024  & 1.44($-$21)   &  3.7      &     & 4.7     \\ 
16 0 15 1 &  1115.177  & 1.35($-$21)   &  1.1      & 2.8 & 6.6     \\ 
36 0 35 1 &  1132.786  & 1.34($-$21)   &  0.9      &     & 4.6     \\ 
37 1 36 0 &  1133.671  & 1.24($-$21)   &  4.0      &     & 4.6     \\ 
15 1 14 0 &  1116.703  & 1.22($-$21)   &  0.8      & 3.1 & 6.8     \\ 
38 0 37 1 &  1134.454  & 1.14($-$21)   &  1.9      &     & 4.5     \\ 
28 1 27 2 &  1123.772  & 1.11($-$21)   &  2.0      & 1.3 & 5.3     \\ 
29 2 28 1 &  1129.009  & 1.10($-$21)   &  1.8      & 1.3 & 5.3     \\ 
30 1 29 2 &  1125.905  & 1.10($-$21)   &  2.3      & 1.1 & 5.2     \\ 
27 2 26 1 &  1127.967  & 1.09($-$21)   &  2.3      & 1.4 & 5.4     \\ 
26 1 25 2 &  1121.562  & 1.09($-$21)   &  4.1      & 1.4 & 5.5     \\ 
14 0 13 1 &  1113.231  & 1.08($-$21)   &  1.8      & 3.3 & 7.0     \\ 
31 2 30 1 &  1130.139  & 1.07($-$21)   &  2.4      &     & 5.1     \\ 
25 2 24 1 &  1127.001  & 1.06($-$21)   &  1.6      & 2.6 & 6.5     \\ 
32 1 31 2 &  1127.953  & 1.05($-$21)   &  0.7      & 0.9 & 5.0     \\ 
39 1 38 0 &  1135.318  & 1.05($-$21)   &  2.4      &     & 4.5     \\ 
24 1 23 2 &  1119.287  & 1.03($-$21)   &  1.6      & 1.6 & 5.6     \\ 
33 2 32 1 &  1131.357  & 1.01($-$21)   &  2.8      & 1.0 & 5.0     \\ 
23 2 22 1 &  1126.090  & 9.75($-$22)   &  1.4      & 1.9 & 5.8     \\ 
34 1 33 2 &  1129.918  & 9.65($-$22)   &  4.5      &     & 4.9     \\ 
13 1 12 0 &  1115.399  & 9.61($-$22)   &  0.6      & 3.6 &         \\ 
40 0 39 1 &  1136.114  & 9.50($-$22)   &  2.6      &     &         \\ 
22 1 21 2 &  1116.967  & 9.44($-$22)   &  1.3      & 1.8 & 5.8     \\ 
35 2 34 1 &  1132.657  & 9.17($-$22)   &  3.1      &     & 4.9     \\ 
21 2 20 1 &  1125.209  & 8.81($-$22)   &  1.2      & 2.2 & 6.0     \\ 
36 1 35 2 &  1131.808  & 8.65($-$22)   &  0.5      &     & 4.9     \\ 
41 1 40 0 &  1136.964  & 8.60($-$22)   &  2.9      &     &         \\ 
20 1 19 2 &  1114.626  & 8.35($-$22)   &  4.6      & 2.1 & 6.0     \\ 
12 0 11 1 &  1111.255  & 8.19($-$22)   &  1.1      & 3.8 &         \\ 
37 2 36 1 &  1134.029  & 8.10($-$22)   &  3.5      &     & 4.8     \\ 
19 2 18 1 &  1124.329  & 7.74($-$22)   &  0.4      & 2.5 & 6.3     \\ 
42 0 41 1 &  1137.768  & 7.74($-$22)   &  1.0      &     & 4.4     \\ 
38 1 37 2 &  1133.632  & 7.52($-$22)   &           &     & 4.8     \\ 
\hline\hline
\end{longtable}

\begin{longtable}{rrrrrr}
\caption{Calculated \oz\ intensities  (PES1 and our DMS) of the (001), (100) and (101) bands compared with 
the measurements of Thomas {\it et al.} \cite{10ThVoDe.O3}.
Measured intensities are given as 100\%\ abundance in units of cm/molecule
with powers of 10 in parenthesis. Our intensities are given
as a percentage difference to these values.}\\
\hline
$\tilde{\nu}$ (\cm) & Intensity  &exp. unc.\%& Assign $E_{up}$ & Assign  $E_{low}$&(o-c)/c \% \\
\hline
 973.0979 & 1.94($-$22) & 2.3 & 58 2 57 (001) & 59 2 58 (000) &	 2.6     \\
 974.5576 & 2.92($-$22) & 2.2 & 57 0 57 (001) & 58 0 58 (000) &	 1.5     \\
 974.5809 & 2.36($-$22) & 1.5 & 57 1 56 (001) & 58 1 57 (000) &	 0.6      \\
 974.5987 & 8.80($-$23) & 3.9 & 55 9 46 (001) & 56 9 47 (000) &	-1.9     \\
 988.8183 & 1.07($-$21) & 0.8 & 47 5 42 (001) & 48 5 43 (000) &	 0.6      \\
 994.5107 & 1.09($-$21) & 1.1 & 39 10 29 (001)& 40 10 30 (000)&	 0.2     \\
1048.6125 & 1.53($-$21) & 1.6 & 19 13 6 (001) & 18 13 5 (000) &	 0.6      \\
1051.3383 & 1.15($-$21) & 0.8 & 27 12 15 (001)& 26 12 14 (000)&             \\
1054.0585 & 1.63($-$21) & 1.6 & 23 12 11 (001)& 22 12 10 (000)&      1.0     \\
1058.0241 & 1.27($-$21) & 1.5 & 36 11 26 (001)& 35 11 25 (000)&      0.0     \\
1059.5267 & 1.99($-$21) & 1.1 & 36 10 27 (001)& 35 10 26 (000)&	 0.1      \\
1065.9838 & 1.63($-$21) & 1.4 & 44 7 38 (001) & 43 7 37 (000) &	-0.1     \\
1066.0126 & 2.14($-$21) & 1.5 & 47 1 46 (001) & 46 1 45 (000) &	-0.8     \\
1070.4355 & 1.83($-$21) & 1.4 & 47 3 44 (001) & 46 3 43 (000) &	 0.4      \\
1071.0834 & 1.71($-$21) & 1.0 & 47 4 43 (001) & 46 4 42 (000) &         \\
1071.9168 & 1.19($-$21) & 0.8 & 49 4 45 (001) & 48 4 44 (000) &  0.0     \\
          &          &  &	              &	              &         \\
1077.9813 & 3.69($-$22) & 1.1 & 33 8 26 (100) & 32 9 23 (000) &	 4.5   \\
1078.8580 & 3.44($-$22) & 1.2 & 30 0 30 (100) & 31 1 31 (000) &	-4.9   \\
1079.4090 & 4.06($-$22) & 0.9 & 20 6 14 (100) & 19 7 13 (000) &	 4.8   \\
1083.0592 & 3.39($-$22) & 1.3 & 17 5 13 (100) & 16 6 10 (000) &	 5.2   \\
1100.0901 & 5.00($-$22) & 0.8 & 22 3 19 (100) & 21 4 18 (000) &  4.5     \\
1101.7616 & 4.93($-$22) & 0.7 & 25 3 23 (100) & 24 4 20 (000) &  4.5     \\  
1104.0767 & 4.94($-$22) & 0.9 & 26 3 23 (100) & 25 4 22 (000) &  4.4     \\  
1109.2699 & 5.71($-$22) & 0.7 & 10 0 10 (100) & 9 1 9 (000)   &  6.7     \\  
1111.5000 & 3.25($-$22) & 1.4 & 7 1 7 (100)   & 6 0 6 (000)   &  8.2     \\  
1115.1773 & 1.32($-$21) & 0.7 & 16 0 16 (100) & 15 1 15 (000) &  4.2      \\ 
1116.7032 & 1.20($-$21) & 1.2 & 15 1 15 (100) & 14 0 14 (000) &  5.3     \\  
1118.0486 & 1.45($-$21) & 1.1 & 17 1 17 (100) & 16 0 16 (000) &  4.6      \\ 
1119.2868 & 1.02($-$21) & 0.8 & 24 1 23 (100) & 23 2 22 (000) &  4.4    \\   
1119.4463 & 1.67($-$21) & 1.0 & 19 1 19 (100) & 18 0 18 (000) &  4.3     \\  
1121.5617 & 1.07($-$21) & 0.9 & 26 1 25 (100) & 25 2 24 (000) &  3.7     \\  
1122.4009 & 1.95($-$21) & 1.3 & 23 1 23 (100) & 22 0 22 (000) &  3.6      \\ 
1122.5439 & 1.98($-$21) & 1.3 & 24 0 24 (100) & 23 1 23 (000) &  3.3     \\  
1123.8503 & 1.51($-$22) & 1.0 & 6 3 3 (100)   & 5 2 4 (000)   &  8.2     \\  
1123.9457 & 2.00($-$21) & 1.2 & 25 1 25 (100) & 24 0 24 (000) &  3.6      \\ 
1124.2947 & 1.99($-$21) & 1.3 & 26 0 26 (100) & 25 1 25 (000) &  3.2    \\   
1124.3286 & 7.61($-$22) & 1.0 & 19 2 18 (100) & 18 1 17 (000) &  4.5     \\  
1126.6751 & 4.60($-$22) &     & 36 2 34 (100) & 35 3 33 (000) &  5.5    \\   
1127.2348 & 2.64($-$22) &     & 10 3 7 (100)  & 9 2 8 (000)   &  8.1      \\ 
1129.2358 & 3.53($-$22) &     & 13 3 11 (100) & 12 2 10 (000) &  7.0     \\  
1129.2765 & 1.37($-$22) &     & 5 4 2 (100)   & 4 3 1 (000)   &  8.9      \\ 
1132.6569 & 9.07($-$22) & 0.5 & 35 2 34(100)  & 34 1 33 (000) &  3.7      \\ 
1132.7860 & 1.33($-$21) & 0.5 & 36 0 36 (100) & 35 1 35 (000) &  3.6     \\  
1132.8114 & 4.83($-$22) & 0.7 & 19 3 17 (100) & 18 2 16 (000) &  5.7     \\  
1133.4335 & 2.65($-$22) & 0.7 & 10 4 6 (100)  & 9 3 7 (000)   &  8.1     \\  
1133.5869 & 1.37($-$22) & 1.2 & 18 2 16 (100) & 17 1 17 (000) &  5.2     \\  
1133.6317 & 7.44($-$22) & 1.1 & 38 1 37 (100) & 37 2 36 (000) &  3.6     \\  
1133.6712 & 1.23($-$21) & 1.0 & 37 1 37 (100) & 36 0 36 (000) &  3.7      \\ 
1133.7245 & 5.04($-$22) & 0.4 & 21 3 19 (100) & 20 2 18 (000) &  4.5     \\  
1133.9786 & 3.02($-$22) & 1.4 & 42 2 40 (100) & 41 3 39 (000) &  3.1      \\ 
1134.0288 & 8.00($-$22) & 1.2 & 37 2 36 (100) & 36 1 35 (000) &  3.4    \\   
1134.2514 & 2.91($-$22) & 0.6 & 11 4 8 (100)  & 10 3 7 (000)  &  7.5     \\  
          &          &  &                  &               &          \\  
2086.1304 & 4.94($-$22) & 0.5 & 19 10 9 (101) & 20 10 10 (000)&  -1.4     \\ 
2086.4723 & 8.82($-$22) & 0.5 & 21 8 13 (101) & 22 8 14 (000) &  -2.1    \\  
2086.9846 & 3.64($-$22) & 0.9 & 18 9 10 (101) & 19 9 11 (000) &  -1.0    \\  
2087.5873 & 9.23($-$22) & 0.8 & 20 8 13 (101) & 21 8 14 (000) &  -1.4     \\ 
2088.2490 & 3.68($-$22) & 0.3 & 17 9 8 (101)  & 18 9 9 (000)  &  -0.8   \\   
2088.6874 & 9.44($-$22) & 0.5 & 19 8 11 (101) & 20 8 12 (000) &  -2.0    \\  
2089.7729 & 9.59($-$22) & 0.7 & 18 8 11 (101) & 19 8 12 (000) &  -2.4   \\   
2103.2106 & 7.43($-$22) & 0.4 & 12 12 1 (101) & 12 12 0 (000) &  -0.7     \\ 
2106.5742 & 8.25($-$22) & 0.2 & 14 8 7 (101)  & 14 8 6 (000)  &  -2.3    \\  
2116.8712 & 4.77($-$22) & 0.7 & 15 10 5 (101) & 14 10 4 (000) &  -1.7     \\ 
2116.9201 & 9.11($-$22) & 0.8 & 11 7 4 (101)  & 10 7 3 (000)  &  -2.8     \\ 
2117.4047 & 5.21($-$22) & 0.7 & 16 10 7 (101) & 15 10 6 (000) &  -1.8    \\  
2117.9200 & 5.54($-$22) & 0.9 & 17 10 7 (101) & 16 10 6 (000) &  -1.7    \\  
2125.2044 & 6.59($-$22) & 0.7 & 26 8 19 (101) & 25 8 18 (000) &  -5.2    \\  
2125.6791 & 6.33($-$22) & 2.0 & 27 8 19 (101) & 26 8 18 (000) &          \\  
2132.7605 & 1.97($-$22) & 0.9 & 44  5 40 (101)& 43  5 39 (000)&  -3.6    \\  
2132.8026 & 2.36($-$22) & 1.0 & 43  5 38 (101)& 42  5 37 (000)&  -1.2     \\ 
\hline\hline
\end{longtable}

\begin{longtable}{rrrrr}
\caption{Calculated  intensities  (PES1 and our DMS) of the \oz\  (003) band compared with 
the HITRAN 2016. 
HITRAN intensities are given as 100\%\ abundance in units of cm/molecule
with powers of 10 in parenthesis. Our intensities are given
as a percentage difference to these values.}\\
\hline
$\tilde{\nu}$  & intensity     &   Assignment     & Assignment    & (o-c)/c \% \\
(\cm)& HITRAN        &    upper         &  lower        &            \\
           &               &                  &               &            \\
\hline
3002.8414&	1.39($-$22)&       32 2 31 (003)&  33 2 32 (000) &     10.1  \\
3007.9106&	1.73($-$22)&	29 3 26 (003)&  30 3 27	(000) &     10.8  \\
3014.1642&	2.41($-$22)&	25 3 22 (003)&  26 3 23	(000) &     10.7  \\
3020.9156&	3.46($-$22)&	21 1 20 (003)&	22 1 21 (000) &     11.5  \\
3022.5631&	3.78($-$22)&	20 1 20 (003)&	21 1 21 (000) &     11.3  \\
3025.3709&	4.04($-$22)&	18 1 18 (003)&	19 1 19 (000) &     11.2  \\
3036.4846&	3.73($-$22)&	 9 0  9 (003)&  10 0 10	(000) &     10.2  \\
3050.7231&	2.50($-$22)&	 7 2  5 (003)&   6 2  4	(000) &	     8.3  \\
3052.3788&	3.14($-$22)&	11 3  8 (003)&  10 3  7	(000) &      7.8  \\
3056.9845&	3.93($-$22)&	19 1 18 (003)&	18 1 17 (000) &      7.8  \\
3057.6913&	3.27($-$22)&	23 1 22 (003)&	22 1 21 (000) &      7.2  \\
3057.7125&	2.46($-$22)&	27 1 26 (003)&	26 1 25 (000) &      6.4  \\

\hline\hline
\end{longtable}

\section{Conclusions}

This work addresses from a theoretical perspective the long 
standing problem of inconsistency  of the IR intensities of ozone.
For the 10 \muu\ band,  for which different laboratory measurements yield
results which differ by up to 4 \%,  our calculations
coincide within 1 \% with the measurements selected by HITRAN
and confirmed recently by simultaneous  microwave and infrared
measurements \cite{17DrCrYu.O3}. Furthermore, our calculations for absorption lines in the 5 \muu\ band 
using the same surfaces gives results within
the experimental accuracy of about 2\%. 
Finally, our results for 3 \muu\ suggest that a significant
lowering of the intensities given by HITRAN in this region
is needed; this result would appear also
to be in line with recent atmospheric observations \cite{toonoz}.

The next step in this study is to produce comprehensive line lists
for \oz\ and its isotopologues, and use them as input to atmospheric
 radiative transfer models to demonstrate that their use does indeed lead to consistent
retrievals. This involves producing multiple line lists for each isotopologue
as a check on line-by-line accuracy and stability \cite{jt522,jt625}. 
Work in this direction is currently in progress.  

The most important absorption of solar radiation by ozone is in the
UV.  The consistency between IR and UV intensity measurements is a
major outstanding problem of ozone spectroscopy.  An important step towards resolving this consistency
problem would be achieved with the accurate (within 1 \%) computation of both IR and UV
bands absorption. For the UV intensity calculation several components
are necessary. A program to compute rovibronic transition intensities
for triatomic molecules has been published recently by one of us
\cite{jt697}.  An accurate ground electronic state PES is an important
part of the electronic spectra calculations; this is presented here.
Accurate \ai\ calculated transition dipole moment surfaces and
electronic excited states PES are available in the literature \cite{10ScMcxx.O3}.           
Fitting of the excited-state PES to 
experimental data and further improvement of the dipole moment
calculations are in progress. This should yield accurate electronic
transition intensities calculations. Again, part of this procedure is
the accurate ground electronic state PES, presented in this work.

We include  as supplementary
material the fitted ozone PES. 
We searched the literature carefully and failed to find any freely available 
ozone PES which
had been accurately fitted to reproduce the known empirical energy levels of the ozone ground electronic state,
though results using such PESs have been presented many times.
Thus, the present PES, given in the supplementary material  as
a Fortran program, is the first freely-available, accurate \oz\ PES for ozone.

\section{Acknowledgment}
We thanks Geoff Toon (JPL) for sharing his assessment  of the HITRAN 2016
ozone line list with us and for helpful discussions.
This work was supported by  the UK Natural Environment
Research Council under grant NE/N001508/1. NFZ, IIM and AAK
acknowledge support by State Project IAP RAS No.0035-2014-009.

\bibliographystyle{elsarticle-num}

\end{document}